
\def\d#1/d#2{ {\partial #1\over\partial #2} }

\newcount\sectno

\newcount\subsectno

\def\subsect{\global\advance\subsectno by1 \the\sectno.\the\subsectno }
 \def\sect{\subsectno=0 \global\advance\sectno by1 \the\sectno }

\def\pdr{\partial}

\def\de{\delta}
\def\eps{\epsilon}
\def\half{{1\over 2}}
\def\tr{\hbox{tr}}

\def\linebreak{\hfil\break}


\newcount\eqnumber
\def\beq{ \global\advance\eqnumber by 1 $$ }
\def\eeq{ \eqno(\the\eqnumber)$$ }
\def\n{\global\advance \eqnumber by 1\eqno(\the\eqnumber)}
\def\puteqno{
\global\advance \eqnumber by 1 (\the\eqnumber)}
\def\beqs{$$\eqalign}
\def\eeqs{$$}


\def\ifundefined#1{\expandafter\ifx\csname
#1\endcsname\relax}

\newcount\refno \refno=0  
\def\[#1]{
\ifundefined{#1}
\advance\refno by 1
\expandafter\edef\csname #1\endcsname{\the\refno}
\fi[\csname #1\endcsname]}
\def\refis#1{\noindent\csname #1\endcsname. }

\def\label#1{
\ifundefined{#1}
\expandafter\edef\csname #1\endcsname{\the\eqnumber}
\else\message{label #1 already in use}
\fi{}}
\def\(#1){(\csname #1\endcsname)}
\def\eqn#1{(\csname #1\endcsname)}

\baselineskip=15pt
\parskip=10pt




\def\sign{\;{\rm sgn}\;}
\magnification=1200

\centerline{\bf Poisson Brackets of Wilson Loops  and
                               Derivations of Free Algebras}

\centerline{S. G. Rajeev$^{1,2}$ and O. T. Turgut
$^{1}$}
\centerline{\it $^1$Department of Physics and Astronomy}
\centerline{\it University of Rochester, Rochester,NY 14627, USA}
 \centerline{\it $^2$ The  Erwin Schrodinger International Institute
 for Mathematical Physics}
\centerline{\it Pasteurgasse 6/7, Vienna, Austria}
\centerline{\it email: rajeev or turgut@urhep.pas.rochester.edu}

\centerline{\bf Abstract}

We describe a finite analogue of the Poisson Algebra of  Wilson
Loops in  Yang--Mills  theory. It is shown that this
algebra arises in an apparently completely different context:
as a  Lie algebra of vector fields on  a
non--commutative  space. This
suggests that non--commutative geometry plays a fundamental role in the
 manifestly gauge invariant formulation of Yang--Mills
theory.  We also construct the   deformation of the loop algebra induced
by quantization, in the large $N_c$ limit.
\vfill\break

{\it Algebras of Wilson Loops}

A central problem of particle physics is to find a formulation of
Yang--Mills theory in terms of gauge invariant variables. There is a
large literature on this subject, starting with pioneering work of
Mandelstam. \[loops]
Such a reformulation of Yang--Mills theory must involve as yet
unknown
geometrical  principles, as the principle of gauge invriance would be
empty. We should  discover these geometrical structures
 by starting with the conventional formalism of gauge theory  and
rewriting it  in terms of gauge invariant variables. A loose analogy
can be made with the process by which symplectic geometry was
discovered to be the foundation of classical mechanics.

In this paper we will
show that the fundamental Poisson brackets of Yang--Mills theory
have an interpretation as an algebra of vector fields in a
non--commutative space, a sort of non--commutative generalization of
the Virasoro algebra. The precise mathematical formulation is
possible only for a finite ( regularized) version of the  theory, but
the ideas extend in a formal way to the continuum theory.

  A natural
choice of  gauge invariant variable  in Yang--Mills theory is
the Wilson loop variable.
It is just the trace of the parallel transport operator around a loop.
We can describe the symplectic structure of
 classical Yang--Mills  theory in terms of Poisson brackets of these
 variables. However, in the usual canonical formalism, where initial
data is given on a space--like surface, this leads to either a
 trivial answer or to an impossibly  complicated one.
 If the loop lies entirely on a
 space--like surface, the Poisson brackets will vanish since the
 components of the gauge field on a space--like surface commute: they
 are like the $q$  variables of classical mechanics. If the loop has
 a finite extend into the time--like direction on the other hand, the
 Poisson brackets cannot be obtained  without solving the equations of
 motion: this is like asking for $\{q(t),q(t')\}$ at unequal times in
 classical mechanics. One way around this impasse is to introduce
 loop
 variables   involving the electric field ( which is the canonical
 conjugate of the Yang--Mills potential)  but this does not have the
 elegance and simplicity of a formalism involving Wilson loop
 variables alone.

We showed in some recent papers \[rajturgut] that the Wilson Loops in classical
Yang--Mills theory which lie  on a null surface satisfy simple
Poisson brackets.( There is also a large literature on the null
cone formalism for gauge theories. See for example,
ref.\[nullcone]. )
In a formalism in which initial data is given on
null surfaces there is thus a natural way of encoding the canonical
structure of Yang--Mills theory in terms of gauge invariant
variables.  In this paper we will show that an analogue of this
Loop algebra arises as  derivations  of the  free algebra
on a finite number of variables. This allows us to construct a Lie
group of which the  finite loop algebra is the Lie algebra. Morover,
we will construct a  symplectic realization analogous to the
realization on the Yang--Mills phase space. We will also obtain a
quantum deformation of this algebra and obtain the contraction
corresponding to the large $N_c$ limit. It is conjectured that this
large $N_c$--limit algebra also has a symplectic  realization, but we
are as yet unable to construct it. This would be of great interest in
Yang--Mills theory, as it would help us  discover the phase space of
gauge invariant observables of that theory.

Let us now describe the situation a little  more
explicitly\[rajturgut].  We will consider pure Yang--Mills theory on
flat Minkowski space, with initial data given on a null cone. The
field will then be determined at all points in the future of the cone
by the Yang--Mills equations. (See Ref. \[rajturgut] for details.)
It will be particularly convenient to choose as initial surface the
null cone at past  time--like infinity, ( called ${\cal I}^-$ in Ref.
\[penrose]) so
that all points not on the cone are  in its future. (This will also restore
spatial translation invariance.)
This can be accomplished by
using a conformally equivalent metric
\beq
     \hat{ds}^2=dU(dU+2dR)-\sin^2 R\;\; q_{ij}(z)dz^idz^j
\eeq
instead of the flat metric on Minkowski space.
(Here, $q_{ij}$ is the standard metric on $S^2$.)
Yang--Mills equations
are conformally invariant in four dimensions,
 so this conformal change of the metric will not change the theory.
${\pdr\over \pdr U}$  is a time--like vector and
${\pdr \over \pdr R}$ is a  null vector. We will regard
${\pdr\over \pdr U}$ as defining the  time direction.
Also, Minkowski space corresponds to the region
\beq
-\pi< U<\pi \quad -\pi< U+2R<\pi
\eeq
The null cone at past  time--like infinity, ${\cal I}^-$ will
be the surface $U=-\pi$.

 Since the Yang--Mills equations  are of first order in the time
 variable $U$,
 initial data consist of prescribing
the value of the gauge potential on ${\cal I}^-$.
We can set  $A_R=0$, by a choice of gauge. Also,  $A_U$
is just a Lagrange multiplier ( its time derivative does not appear
in the Lagrngian) so that  the dynamical variables are  the
transverse components $A_i$. The main simplification of the null
formalism is that these variables are in a sense canonically
conjugate
 to each other with equal time Poisson brackets
\beq
\{A_{ib}^a(z,R),A_{jd}^c(z',R')\}=\half\delta^a_d\delta^c_b
                               q_{ij}(z)\delta(z-z')\sign(R-R').
\eeq
Here $a,b=1,2,\cdots N_c$.  We will consider the gauge group to be
 $U(N_c)$ so that the matrices $A_i$
are hermitian.
These Poisson  brackets follow from the Yang--Mills action by
a straight forward application of the canonical formalism.

Define  thus the inverse of the
symplectic form on the phase
space of Yang--Mills theory:
\beq
     \omega^{ab}_{ijcd}(z,R,z',R')=
     \delta^a_d\delta^b_c\omega_{ij}(z,R,z',R')
\eeq
where,
\beq
     \omega_{ij}(z,R,z',R')=\half q_{ij}(z)\delta(z-z')\sign(R-R').
\eeq

Now let $\xi:S^1\to {\cal I}^-$  be  a closed curve on the light cone.
Given a Yang--Mills field $A$, we can define a complex valued
function $W[\xi]$  on the space of closed curves, the trace of the
parallel transport operator (holonomy)  in the basic $N_c$
dimensional
representation.  In the case of $U(N_c)$,
this loop variable is complex valued, but it satisfies the condition
\beq
     W[\bar\xi]=W^*[\xi]
\eeq
where $\bar\xi $ is the curve $\xi$ with the opposite orientation.

This $W[\xi]$ is the `Wilson Loop' associated to this
Yang--Mills configuration. For each $\xi$, $W[\xi]$ is a function on
the Yang--Mills phase space so that it is possible to compute  the
Poisson bracket of a pair of loop variables. We will get,
\beqs{
     \{W[\xi],W[\tilde\xi]\}&=\int ds dt
    \dot\xi^i(s){\dot{\tilde{\xi}}}^j(t)
\omega_{ij}(\xi(s),\tilde\xi(t))W[\xi\circ_{st}\tilde\xi]. \cr
}\eeqs
Due to the delta function in the symplectic form, only points $s,t$ where
$\xi^i(s)=\tilde\xi^i(t)$  will contribute to the integral;
 i.e., the projections of
the curves to the sphere must intersect at parameter value $s$ for the
first curve and $t$ for the second.  There will be a null line
segment joining the points  $\xi(s)$ and $\tilde\xi(t)$ for this intersection.
In this case,
$\xi\circ_{st}\tilde\xi$ is the product curve, defined as follows:
describe the curve $\xi$  starting  and ending at $s$; jump to the point
$\tilde\xi(t)$ along the null line segment; describe the curve
$\tilde\xi$ starting  and ending at $\tilde \xi(t)$;
 jump back to $\xi(s)$ along the null
segment. Thus the  product  is also a closed curve. The pieces of the curve
along
null lines will not contribute to $W[\xi\circ_{st}\tilde\xi]$ since
we have chosen a gauge   where
the null component of the gauge field vanishes.  Also, there will be
generically only a finite number of intersection points, so that the
integral on the right hand side can be actually evaluated to yield a finite
sum.
We will not need the explicit expression, which is given in  Ref.
\[rajturgut]. This last property depends on the dimension of
space--time being four; in higher than four dimensions  generically
there are no intersections while in lower dimensions there is a
continuum of such intersections.

The Poisson brackets above could have been motivated based purely on
the geometry of loops on the null cone. Causality requires that loops
on a null cone which have no  intersections when projected to the
space--like surface, must  commute. This explains the delta function
in $\omega(z,R,z',R')$. The factor $\sign(R-R')$ is also natural, as
it simply keeps track of which event is to the future along the null
direction, and makes the bracket anti--symmetric.

Indeed, geometrically, these are  the most natural
definitions possible for $\omega$  as well as the product of the
loops. Note that the bracket is invariant under the change of
parametrization of the loop. In fact the rhs will only depend on
$q_{ij}$ only through the
angle of intersection of the tangent vectors $\dot \xi(s),
\dot{\tilde\xi}(t)$; thus the algebra is invariant under conformal
transformation of the metric on the two sphere. But every metric on
the two sphere is conformal to the standard one, so we see that the
algebra is in fact independent of the choice of metric also.

Once the Poisson brackets are postulated,
the Jacobi identity  can be proved directly. The
Yang--Mills phase space then arises as the solutions of some
algebraic constraints satisfied by the loop variables, due to
Mandelstam. We proposed that these constraints be viewed as
describing the co--adjoint orbits of the above Lie algebra. In this
way, Yang--Mills theories with different Unitary
groups as gauge groups, would arise as different realizations of
the same universal Lie algebra.     For more details we refer  the
reader to
Ref. \[rajturgut]

We would like to understand the above Lie  algebra of loops  better.
In particular we
would like to have a finite analogue which can be studied by
more rigorous methods; also  it would be good to have a different
 situation in which this loop algebra  arises so that we can have
a  point of view to Yang--Mills theory not based on the gauge group.
 Another natural  object
to study is the group associated to the above Loop algebra. Finally
it is very important to understand the quantum  deformations of this algebra
and its large $N_c$-limit.
In this paper we will in
fact arrive at a finite analogue of the loop algebras and their
groups, starting from
 considerations quite different from Yang--Mills theories; i.e., the
 derivations and  automorphisms of Free algebras. We will also
 construct a quantum deformation and its large $N_c$ limit.

Another situation where Poisson  brackets of Wilson loops appear is in
Chern--Simons theory.  There also the spatial  components of the
gauge field are canonically conjugate to each other; the Wilson loops
     on a space-like 2--surface satisfy the above algebra except that
     \beq
     \omega_{i,j}(z,z')=\eps_{ij}(z)\delta^2(z-z')
\eeq
where $\eps_{ij}(z)dz^i\wedge dz^j$  is the volume form on the space--like
 surface.
The product of loops realtive to a pair of co-incident points is
defined  as before as one loop followed by the other.Thus our
considerations should also be of interest in the context of
topological field theories.

{\it The Free Algebra and its Automorphisms}\hfill\break

Let ${\cal T}_M$  be  the real free algebra on $M$ variables. \[free] It is a
graded vector space, the part of order $m$ ( for $m=0,1,2\cdots$)
being just the set of all tensors of type $(0,m)$ on an $M$
dimensional real vector space. Note that no symmetry  of any kind is
required on these tensors. The multiplication rule on the  algebra is
defined by the direct ( or tensor) product). ${\cal T}_M$ is a
non--commutative  but associative algebra with identity.

More explicitly,introduce variables $\xi^i$ for $i=1,\cdots M$
satisfying no relations whatever. A typical element of ${\cal T}_M$
is  a polynomial in these variables,
\beq
     T(\xi)=\sum_{m=0}^{\infty}T_{i_1i_2\cdots i_m}\xi^{i_1}\xi^{i_2}
     \cdots \xi^{i_m}.
\eeq
$T_{i_1i_2\cdots i_m}$  are the components with respect to  some
basis in $R^M$ of a tensor
$T$ of type $(0,m)$.   Since $T(\xi)$ is assumed to be a polynomial,
only a finite number of terms on the right hand side of the above series are
non--zero: only a finite number of the tensors $T_{i_1\cdots i_m}$
are non--zero.

In this language, multiplication is defined  as follows
\beq
  (ST)(\xi)=\sum_{m,n=0}S_{i_1\cdots i_m}T_{j_1\cdots j_n}
          \xi^{i_1}\cdots\xi^{i_m}\xi^{j_1}\cdots \xi^{j_n}.
\eeq
There is no problem with convergence  of the series since  only a
finite number of terms are non--zero.  In fact this comment applies
to almost all the formally infinite series below. ( The exception is
where we speak of inverting a transformation of the variables.)

If these variables $\xi$ had commuted with each other, the algebra
would just have been the commutative algebra of functions
(polynomials) on $R^M$.
The tensors
would all have been symmetric and multiplication would have been
the symmetrized  tensor product.  This algebra has as automorphisms
the group of diffeomorphisms of $R^M$. (Actually a diffeomorphism
 will in general map a polynomial to an infinite series, so we will
really need  ${\cal T}_M$ to extend to an appropriate topological
vector space  to make this  possible.)
Infinitesimally, this would correspond to the Lie algebra of vector
fields, whose components are polynomials, which  form the  derivations of
the commutative  algebra.

Thus we can regard ${\cal T}_M$ as  the set of `functions' on a
non--commutative space in the spirit of non--commutative geometry \[connes].
This is perhaps  the most non--commutative case in the sense that the
co--ordinates satisfy no relations at all. Now let us determine the Lie
algebra of derivations, ${\cal V}_M$ which will be
the non--commutative analogue
of the algebra of vector fields.  A derivation $v$ is determined by
its effect on the generators:
\beq
     v(\xi)^i=\sum_{m=1}v^i_{i_1\cdots i_m}\xi^{i_1}\cdots \xi^{i_m}
\eeq
where  it is assumed that  only a finite number of terms in the sum are
non--zero .  The effect of $v$  on an arbitrary element of
${\cal T}_M$ is given by the Leibnitz rule:
\beq
     v(T)(\xi)=\sum_{m,n=1}^\infty\sum_{k=1}^m
     T_{i_1\cdots i_m}v^{i_k}_{j_1\cdots j_n}\xi^{i_1}\cdots \xi^{i_{k-1}}
\xi^{j_1}\cdots \xi^{j_n}\xi^{i_{k+1}}\cdots \xi^{i_m}
\eeq
A basis (analogous to the Weyl basis for $gl(M)$) for ${\cal V}_M$
 is given by the
elements $E_i^{i_1\cdots i_m}$ defined by
\beq
     E_i^{i_1\cdots i_m}(\xi)^j=\delta^j_i\xi^{i_1}\cdots \xi^{i_m}.
\eeq
In the commutative case they correspond to the vector fields
$\xi^{i_1}\cdots \xi^{i_m}{\pdr\over \pdr {\xi^i}}$.
They satisfy the commutation relations,
\beqs{
     [E_i^{i_1\cdots i_m}, E_j^{j_1\cdots j_n}]&=
\sum_{l=1}^n \delta_i^{j_l}E_j^{j_1\cdots j_{l-1}i_1\cdots i_m
j_{l+1}\cdots j_n}\cr
& -\sum_{k=1}^m \delta_j^{i_k}E_i^{i_1\cdots i_{k-1}j_1\cdots j_n
i_{k+1}\cdots i_m}.\cr
}\eeqs

In the special case $M=1$, all the non--commutativity dissapears, and
${\cal V}_1$ is just the algebra of polynomial vector fields on the
real line.
 Since all the indices must take the value one,
there is just one generator with $m$ superscripts. Suppose we
call it $L_{m-1}$ for $m=0,1,\cdots$. Then the above commutation relation
becomes
\beq
     [L_m,L_n]=(n-m)L_{m+n}\;\;{\rm for}\;\; m=-1,0,1,2,\cdots.
\eeq
This is just the  subalgebra of the Virasoro algebra on which the
central term vanishes. Thus our algebras are, in a sense,
generalizations of this familiar algebra.

Now let $g_{ij}$ be a symmetric positive tensor on $R^M$ and define
${\cal V}^-_M$ to be the subalgebra of tensors that preserve the element
$g(\xi)=|\xi|^2=g_{ij}\xi^i\xi^j$.  In the commutative case these are all the
vector fields tangential to the spheres centered at the origin; these
preserve the distance function $|\xi^2|$. The simplest among these are the
rotations.  In the case of a Free algebra,  we can  see  easily that the
algebra ${\cal V}^-_M$  consists of the set of all elements of the form
\beq
     v^i_{i_1\cdots i_m}=g^{ii_0}w_{i_0i_1\cdots i_m}
\eeq
where $w_{i_0i_1\cdots i_m}$ is a {\it cyclically anti--symmetric} tensor.
Of course such tensors exist only when $m$ is  odd. There is a basis
for ${\cal V}^-_M$,
\beq
     G^{i_0\cdots i_m}=\sum_{k=0}^m (-1)^{k} g^{i_k
     j}E_{j}^{i_{k+1}\cdots i_0i_m\cdots i_{k-1}}
\eeq
in which  the Lie brackets become
\beq
     [G^{i_0\dots i_m},
     G^{j_0\cdots j_n}]=\sum_{k,l=0}^{k=m,l=n}(-1)^{k+l+1}
          g^{i_kj_l}G^{i_{k+1}\cdots i_mi_1\cdots i_{k-1}
               j_{l+1}\cdots j_nj_1\cdots j_{l-1} }.
\eeq

In an exactly analogous fashion, let $\omega$ be an anti--symmetric
non-degenerate tensor. (Clearly this exists only if $M$ is even,
which will be assumed in the following.)
This defines an element
\beq
     \omega(\xi)=   \omega_{ij}\xi^i\xi^j.
\eeq
This is a symplectic analogue of the distance function. This would
have vanished identically in the commutative case.

The subalgebra ${\cal V}^+_M$  which preserves $\omega(\xi)$  is just
the set of  elements such that
\beq
     v^i_{i_1\cdots i_m}=\omega^{ii_0}w_{i_0i_1\cdots i_m}
\eeq
where $w_{i_0i_1\cdots i_m}$ is a {\it cyclically symmetric} tensor.
 There is a basis
for ${\cal V}^+_M$,
\beq
     F^{i_1\cdots i_m}=\sum_{k=1}^m  \omega^{i_k
     j}E_{j}^{i_{k+1}\cdots i_1i_m\cdots i_{k-1}}
\eeq
in which the Lie brackets become
\beq
     [F^{i_1\dots i_m}, F^{j_1\cdots j_n}]=\sum_{k,l=1}^{k=m,l=n}
          \omega^{i_kj_l}F^{i_{k+1}\cdots i_mi_1\cdots i_{k-1}
               j_{l+1}\cdots j_nj_1\cdots j_{l-1} }.
\eeq

Now we will show that this algebra is just a finite version of the
loop algebra we found for Wilson loops. Let  us think of the index
$I={i_1\cdots i_m}$  on the $F^I$  variable as a map
$I:Z_m\to
\{1,\cdots M\}$. Due to cyclic symmetry, this can be viewed
as a `loop'  from the cyclic permutation group $Z_m$ ( which is a
discrete model for the circle) to a space which contains just a
finite number $M$ of points. The product of two loops at point $k,l$
is defined as the loop $I$ starting at $i_{k+1}$  and ending at
$i_{k-1}$ followed by  the loop $J$  starting at $j_{l+1}$  and ending
at $j_{l-1}$:
\beq
     I\circ_{kl}J={i_{k+1}\cdots i_mi_1\cdots i_{k-1}
               j_{l+1}\cdots j_nj_1\cdots j_{l-1} }.
\eeq
This is just the discrete analogue of the product we introduced
earlier.
The  commutation relations of the Lie algebra ${\cal V}^+_M$ are
then,
\beq
     [F^I,F^J]=\sum_{kl}\omega^{i_kj_l}F^{I\circ_{kl}J}.
\eeq
The Wilson loop algebra can  be understood as the
limiting case where  the finite set $Z_m$  is replaced by $S^1$ and
the set $\{1,2,\cdots ,M\}$ is replaced by the light cone
$S^2\times R^+$.  Thus by studying
the algebra ${\cal V}^+_M$  we are studying a  finite model for the
algebra of Wilson loops. The algebras ${\cal V}_M, {\cal V}^-_M,
{\cal V}^+_M$ are all graded Lie algebras. The point is  that the
Lie
bracket of a tensor with  $m$ indices  and one of with  $n$  indices
 has
$m+n-2$ indices.  Thus if we assign a grade of $m-1$ to the space of
tensors  wit $m$ indices, we have a graded Lie algebra. The range of
the grading is $-1,0,1,2$;  the space of
grade $-1$, if nonempty, is a subalgebra with vanishing brackets.

 Although we have
introduced ${\cal V}^+_M$ as a complex Lie algebra, its unitary form,
obtained by imposing
\beq
     v_I^*=v_{\bar I}
\eeq
on the coefficients of an element $v=\sum v_I F^I$,
will be of particular interest. To save on notation we will call this
real Lie algebra also ${\cal V}^+_M$
Here,
\beq
     \bar I=i_m i_{m-1}\cdots i_1
\eeq
is the loop with the opposite orientation. The  conjugation
$v_I\to v^*_{\bar I}$
is an antilinear involution of the complex algebra, therefore
it makes sense to talk about its unitary form.

The algebra ${\cal V}^-_M$ is the finite analogue of the
commutation relations of the Wilson loop in
super--symmetric Yang--Mills  theory.  In the case of supersymmetric
QCD in two dimensions, for example, the bosonic components of the
gauge field can be removed by gauge fixing, and the analogues of the
Wilson loop variables involve only the fermionic fields in
the adjoint represetation.  Thus it is of equal interest to study
${\cal V}^-_M$; we can  develop the two cases in parallel;
 but mostly
 we will speak of  ${\cal V}^+_M$.

{\it Automorphism groups $G^{\pm}_M$}\hfill\break

It is now possible to understand the Lie groups of which
${\cal V}^\pm_M$ are Lie algebras. Thus, we will solve in a finite
context the problem of exponentiating the Lie algebra of Wilson
loops.

First of all, let us consider a group of which ${\cal V}_M$ is the
Lie algebra.    Consider the vector space of tensors of type $(1,m)$  for
$m=0,1,\cdots$. In terms of the variables $\xi^i$, a typical element
would be
\beq
     \phi^i(\xi)=\sum_{m=1}^\infty
     \phi^i_{i_1\cdots i_m}\xi^{i_1}\cdots \xi^{i_m}.
\eeq
Define the composition law  of such functions of $\xi$ in the obvious
way:
\beq
     (\tilde \phi\circ \phi)^i(\xi)=\sum_{m=0}^\infty
     \tilde\phi^i_{i_1\cdots i_m}\phi(\xi)^{i_1}\cdots \phi(\xi)^{i_m} .
\eeq
This  operation is clearly associative and has  identity.

 If we now
restrict to the subset of functions $\phi$ such that the first tensor
in the series above, is invertible,
\beq
     G_M=\{\phi|\det \phi^i_j
\neq 0\}
\eeq
we have a group under the above compostion law. To see this, we note
that given any such $\phi$, unique inverse $\psi$ can be constructed
solving the equation
\beq
     \psi^i(\phi)(\xi)=\xi^i
\eeq
recursively:
\beqs{
     \psi^i_j\ \phi^j_k&=\delta^i_k,\cr
      \psi^i_j\phi^j_{j_1j_2}+
\psi^i_{i_1i_2}\phi^{i_1}_{j_1}\phi^{i_2}_{j_2}&=0,\cr
}\eeqs
etc.The term of order $m$ will determine $\psi^i_{i_1\cdots i_m}$ in terms
of lower
order components of $\psi$ thus establishing the existence  and
uniqueness of an
inverse. In general $\psi$ will have an infinite number of non--zero
terms even when $\phi$  is a polynomial.
\footnote{$^1$} {We must  enlarge our space of allowed transformations to
include
infinite series, in order to be able to  define an inverse. We dont address the
issue of convergence of these series, although it should be possible
to define an appropriate topology on the space of such series with
respect to which $G_M$ is a Lie group.The Lie algebra of $G_M$ will
in fact be the completion of our polynomial derivations ${\cal V}_M$ in  such a
topology.}
This is an algebraic analogue of the inverse function theorem:
$\phi^i_j$  is the analougue of the  derivative at the origin of the function
$\phi^i(\xi)$, so that if it is invertible, we should expect $\phi$  to
be invertible at least locally.

Thus $G_M$ is a group under the above composition law;
 by infinitesimalizing the  composition law we see that this
group  has as  Lie algebra ${\cal V}_M$. We see that $G_M$ is a
non--commutative  analogue of the  diffeomorphism group of $R^M$.

Now it is clear that  groups of which the Lie algebras are
 ${\cal V}^{\pm}_M$
may be defined as below:
\beqs{
     G^-_M&=\{\phi|\det \phi^i_j\neq 0;
     g_{ij}\phi^i(\xi)\phi^j(\xi)=g_{ij}\xi^i\xi^j\}\cr
     G^+_M&=\{\phi|\det \phi^i_j\neq 0;
     \omega_{ij}\phi^i(\xi)\phi^j(\xi)=\omega_{ij}\xi^i\xi^j\}\cr
}\eeqs
which are just the conditions for the distance functions to
be invariant.

{\it Symplectic Realizations}\hfill\break

It would obviously be interesting to look at representations of the
above loop algebras. This should be interesting for example in the
quantum Yang--Mills theory. However, it is quite possible that the
relevant algebras are different in the quantum theory: quantization
could deform the  algebra itself. Therefore we study  first the
classical analogue of a representation, a realization of the Lie
algebra ${\cal V}^+_M$ in terms of  Poisson brackets of some functions
 on symplectic space.

Let $\eta^{ia}_b$  be a set of complex variables  satisfying the
hermiticity condition
\beq
          \eta^{ia*}_b=\eta^{ib}_a.
\eeq

Here,  $i=1,\cdots M$
and $a,b=1,\cdots N_c$  for some positive integer $N_c$. We will
consider only
the case of even $M$. (The indices
$a,b$  will be called color indices, since we will  soon  see an
analogy to Yang--Mills theory. )
       Now impose the Poisson brackets
\beq
     \{\eta^{ia}_b,\eta^{jc}_d\}=\omega^{ij}\delta^a_d\delta^c_b
\eeq
Thus we are just  considering the  real
vector space $R^{MN_c^2}$ with a  symplectic  form that is invariant
under the adjoint action of $U(N_c)$. Now consider the space
 of polynomials invariant under the adjoint action of $U(N_c)$.
A  basis for this space  is  labelled by a discrete loop
$I:Z_m\to \{1,2,\cdots M\}$:
\beq
     f^I(\eta)=\tr\; \eta^{i_1}\eta^{i_2} \cdots \eta^{i_m}
\eeq
The cyclic symmetry of the trace assures us that $f^I$ is independent
of the starting point of the loop $I$.   Moreover,
\beq
     f^{I*}=f^{\bar I}
\eeq
This implies that the coefficients $a_I$ of
an element $a=\sum a_I f^I$ are complex numbers satisfying  $a^*_I=a_{\bar I}$.

Now it is a simple matter to
verify that the Poisson brackets of these functions provide a
realization of the Lie algebra ${\cal V}^+_M$:
\beq
     \{f^I,f^J\}=\sum_{kl}\omega^{i_k,j_l}f^{I\circ_{k,l}J}.
\eeq
The analogy of this realization with the Poisson brackets of the
Wilson loops is obvious.

At the level of the group,       we also have an action of the group
on invariant polynomials of the variables $\eta^i$ by a sort of
`pull--back':
\beq
     \phi^*(h)(\eta)=h(\psi(\eta))
\eeq
where $\psi$ is the inverse of $\phi $  and
\beq
     [\psi(\eta)]^i =\sum_{m=1}\psi^i_{i_1 \cdots i_m}
     \eta^{i_1}\cdots \eta^{i_m}
\eeq
matrix multiplication being implied on the right hand side. If  we restrict to
the sub--group $G^+_M$ the matrix valued function
$\omega_{ij}\eta^i\eta^j$  is invariant under this action.

Clearly if the number of `colors' $N_c$ is  one, the realization
described above has a large kernel.  The variables $\eta^i$ then
satisfy       the relation
\beq
     \eta^i\eta^j-\eta^j\eta^i=0
\eeq
since they commute. The functions $f^I$ then satisfy  the `Mandelstam
identity'
\beq
     f^{I\circ_{kl}J}-f^If^J=0
\eeq
relative to any way of multiplying the two loops $I$ and $J$ at
points $k,l$. More generally, there will be an identity that says
that the anti--symmetric part in $N_c+1$  indices is zero; these are
the finite analogues  of the Mandelstam identities.
For simplicity, let us state $N_c=2$ case;
\beq
     f^{I_1 \circ I_2 \circ I_3}+f^{I_1 \circ I_3 \circ I_1 }
   -f^{I_1 \circ I_2}f^{I_3}-f^{I_1 \circ I_3}f^{I_2}
    -f^{I_2 \circ I_3}f^{I_1} +f^{I_1}f^{I_2}f^{I_3}=0.
\eeq
Here, $I_1$ actually denotes the set $i_1,i_2,....i_{k_1}$, $I_2$ denotes
$j_1,...j_{k_2}$ and $I_3$ refers to $l_1,....l_{k_3}$. Circles are the
products we introduced which corresponds to combining the corresponding
sequences.
Similarly, one can see that writing the all possible antisymmetric
combinations and taking the trace, we get a relation satisfied by
$N_c+1$ generators of the representation. This gives us combinations of
generators with all possible permutations multiplied with the appropriate
sign of the permutation. If we take the cycle decomposition of a permutation
$\pi$ of $N_c+1$ numbers, and denote each cycle as $\pi_k$,  we can
write the result as
\beq
       \sum_{\pi} (-1)^{\pi} f^{I_{\pi_1}}f^{I_{\pi_2}}... f^{I_{\pi_s}}=0
\eeq
where, we used a short hand $f^{I_{\pi_k}}$ to denote $f^{I_{l_{r_{k-1}}}
\circ...\circ I_{l_{r_k}}}$. Here the length of the cycle $\pi_k$
is given by $r_k-r_{k-1}$ and circles again correspond to products.

  These are
precisely the analogues of the  identities satisfied by the Wilson
loop for finite $N_c$( See Ref.\[rajturgut]).
 They simply describe the  fact that $f^I$ is
the trace of an $N_c\times N_c$ matrix. As $N_c\to \infty$ these
identities should dissappear which must be  a reason for the simplicity of
the large $N_c$ limit.

We remark that if we introduce Grassmann variables  $\psi^{ia}_b$
which anti--commute,
\beq
     \psi^{ia}_b\psi^{jc}_d+\psi^{jc}_d\psi^{ia}_b=0
\eeq
and satisfy the super--Poisson bracket
\beq
     \{\psi^{ia}_b, \psi^{jc}_d\}=g^{ij}\de^a_d\de^c_b
\eeq
we also have a super--symplectic realization of ${\cal V}^-_M$:
\beq
     G^{I}\mapsto \tr \psi^{i_1}\cdots\psi^{i_m}.
\eeq
Clearly these $G^I$ are cyclically anti--symmetric and a short
computation will show that their super--Poisson brackets form
a realization of ${\cal V}^+_M$.
By replacing the above Grassmann algebra by a Clifford algebra
\beq
     \psi^{ia}_b*\psi^{jc}_d+\psi^{jc}_d*\psi^{ia}_b=
                         \hbar g^{ij}\de^a_d\de^c_b
\eeq
we also have a quantum
deformation for ${\cal V}^+_M$, analogous to the one in the next
section.

{\it Free Orthogonal Algebra}

Let us consider the $SO(N_c)$ Yang--Mills theory, and obtain a similar
formalism of loops. If we think of $SO(N_c)$ as the real part of $U(N_c)$,
then Wilson loops satisfy,
\beq
     W[\xi]=W[\bar \xi] \quad W^*[\xi]=W[\xi]
\eeq
The loop algebra should also reflect this symmetry, therefore for
$SO(N_c)$ Yang-Mills theory, we obtain the  result:
\beqs{
     & \{ W[\xi_1], W[\xi_2] \}=\cr
 &\int ds dt \bigg(
\dot \xi_1^i(s) \dot \xi_2^j(t)\omega_{ij}(\xi_1(s),\xi_2(t))
  W[\xi_1 \circ_{st} \xi_2] +
   \dot \xi_1^i(s) \dot {\bar \xi_2}^j(t)\omega_{ij}(\xi_1(s),\bar \xi_2(t))
  W[\xi_1 \circ_{st} \bar \xi_2]\bigg).\cr
}\eeqs
One can see that the algebra is invariant under $\xi \mapsto \bar \xi$
\[rajturgut].

 We will see that  the real subalgebra of the unitary
algebra is in fact the above  algebra of Wilson loops.
The real form  can be obtained by imposing the  conditions,
\beq
    v_I=v_{\bar I}\quad v^*_I=v_I
\eeq
for the coefficients of an element $v=\sum v_I F^I$.
One can check that a basis for them is given by
\beq
   H^I=F^I+F^{\bar I}  .
\eeq
We can now calculate the commutator of these basis elements;
\beq
  [H^I,H^J]=\sum_{k,l} \omega^{i_kj_l}(F^{I \circ_{kl} J}+
  F^{I \circ_{kl} {\bar J}}+F^{{\bar I}\circ_{kl} J}+F^{{\bar I} \circ_{kl}
  {\bar J}})
,\eeq
using the previous result on the commutators of $F^I$'s.
Due to the cyclic symmetry we can show that $F^{{\bar I} \circ_{kl} {\bar J}}=
F^{\overline {I \circ_{kl} J}}$. Thus the above expression can be reorganized
as
\beq
 [H^I, H^J]=\sum_{kl} \omega^{i_kj_l}(H^{I \circ_{kl} J} + H^{I \circ_{kl}
   {\bar J}}).
\eeq

This also shows explicitly that the above set of elements constitute a
subalgebra.
We will also briefly describe a symplectic realization of the above algebra.
Let us consider the real symmetric matrices $\eta^{ia}_b$,
where $i=1,2...M$ and $a,b=1,2...N_c$, with the Poisson
bracket,
\beq
    \{ \eta^{ia}_b, \eta^{jc}_d \}=\omega^{ij}(\delta^a_d\delta^c_b+
         \delta^{ac}\delta_{bd})
.\eeq
If we define the $SO(N_c)$ invariant polynomials,
\beq
    h^I={\rm tr}\eta^{i_1}\eta^{i_2}...\eta^{i_n}
\eeq
they provide a symplectic realization of the above algebra. We have
the condition $h^I=h^{\bar I}$ and the Mandelstam constraints as the
kernel of this realization.

It is also possible to discuss the deformation of this symplectic realization
but we will only consider the more physically relevant case of unitary
algebras. The  derivation given below can be extended to the orthogonal case.

{\it Quantum deformation}

It is interesting to see what happens to the above realization  upon
quantization. One approach to quantization is the deformation of the
commutative product of the functions of $\eta^{ia}_b$ by the
so-called  Moyal
product:
\beq
f*g(\eta)=\bigg[e^{-i{\hbar\over 2}\omega^{ij}
             {\pdr\over \pdr\eta^{ia}_b}{\pdr\over \pdr\eta^{'jb}_a} }
               f(\eta)g(\eta')\bigg]_{\eta=\eta'}.
\eeq
(This particular definition of the product  corresponds to Weyl
ordering.\[hormander])
If we apply this multiplication rule
 to the  $U(N_c)$ --invariant polynomials $f^I$, we
will get    a non--commutative associative algebra. The commutator of
this multiplication defines a Lie algebra, which is a quantum
 deformation of our loop algebra ${\cal V}^-_M$. To first order in
 $\hbar$ this commutator is just the Poisson bracket,  so that in
 this limit we recover the previous algebra as a contraction of the
 quantum algebra.      But the general answer is quite formidable,
 at each order $r$ in $\hbar$, there will be terms involving upto $r$
 products of loops.

On the other hand, it is to be expected that some simplifications
will occur in the limit as $N_c\to \infty$.  The point is that the
leading contribution will come from terms where there are the largest
number of possible independent traces, so that we must keep the
terms with the largest number of loops. All the other  terms are
sub--leading order. Neverthless, it turns out that there is  such a term
of leading order in ${1\over N^2_c}$ at each order in $\hbar$; the
limit $N_c\to \infty$ is quite different from the limit
$\hbar \to 0$. But this is also a `classical' limit in that the
commutators of color invariant  observables is of order ${1\over N^2_c}$,
so that they become simultaneously measurable in the limit
$N_c\to \infty$. It is of utmost importance to understand the large
$N_c$ limit of gauge theories; our duscussion identifies the
canonical structure (Poisson brackets of loop variables) of color
singlet observables in the large $N_c$ limit.

Let us now calculate the deformed brackets more explicitly. First of
all note that
\beq
     {\pdr f^I\over \pdr \eta^{ka}_b}=0
\eeq
unless $k$  is equal to one of the elements of the loop
$\{i_1,i_2\cdots i_m\}$. In the case
 $k=i_\mu$  for some $\mu=1,2,\cdots m$,
\beq
     {\pdr f^I\over \pdr \eta^{i_\mu a}_b}=
\bigg[\eta^{i_{\mu+1}}\eta^{i_{\mu+2}}\cdots
          \eta^{i_m}\eta^{i_1}\cdots \eta^{i_{\mu-1}}\bigg]^b_a.
\eeq
Thus differentiation with respect to $\eta^{i_\mu}$
 cuts the loop at  the point with parameter value
$\mu$.

 More generally,
\beq
     {\pdr^r f^I\over \pdr \eta^{k_1a_1}_{b_1}\cdots \pdr \eta^{k_r
     a_r}_{b_r}}=0
\eeq
unless the set $\{k_1,k_2\cdots k_r\}$ is a subset of the set
$\{i_1,i_2\cdots i_m\}$.  Suppose
$\{\mu_1,\mu_2\cdots \mu_r\}\subset\{1,2,\cdots m\} $ and moreover
that $\mu_1<\mu_2\cdots< \mu_r$. Then we can see  that
\beqs{
     {\pdr^r f^I\over \pdr \eta^{i_{\mu_1}
     a_1}_{b_1}\cdots \pdr \eta^{i_{\mu_r} a_r}_{b_r}}
&=
\bigg[\eta^{i_{\mu_1+1}}\eta^{i_{\mu_1+2}}\cdots
                          \eta^{i_{\mu_2-1}}\bigg]^{b_1}_{a_1} \cr
\bigg[\eta^{i_{\mu_2+1}}\eta^{i_{\mu_2+2}}\cdots
                          \eta^{i_{\mu_3-1}}\bigg]^{b_2}_{a_2} \cr
\cdots \bigg[\eta^{i_{\mu_r+1}}\eta^{i_{\mu_r+2}}\cdots
                          \eta^{i_{\mu_1-1}}\bigg]^{b_r}_{a_r} \cr
}\eeqs
 which corresponds to cutting the loop at points
 $\mu_1,\mu_2,\cdots \mu_r $.     It is clearly convenient to
 introduce the matrix,  for $\mu_1<\mu_2\in\{1,2,\cdots m\}$
\beq
     P^b_a(I(\mu_1,\mu_2))=\bigg[\eta^{i_{\mu_1+1}}\eta^{i_{\mu_1+2}}\cdots
                          \eta^{i_{\mu_2-1}}\bigg]^{b}_{a}
\eeq
which represents the parallel transport operator for the  piece of the
loop $I$ from $\mu_1$ to $\mu_2$.
Then, for $\mu_1<\mu_2\cdots <\mu_r$,
\beqs{
     {\pdr^r f^I\over \pdr \eta^{i_{\mu_1}
     a_1}_{b_1}\cdots \pdr \eta^{i_{\mu_r} a_r}_{b_r}}
&=  P^{b_1}_{a_1}(I(\mu_1,\mu_2))P^{b_2}_{a_2}(I(\mu_2,\mu_3))\cdots
 P^{b_r}_{a_r}(I(\mu_r,\mu_1)).
}\eeqs

Now let us consider the general term in the definition of the
deformed product of color invariant functions of the $\eta$:
\beqs{
     f^I*f^J&=f^If^J+\sum_{r=1}^{\infty}
    {1\over r!} \bigg(-{i\hbar\over 2}\bigg)^r
\omega^{i_{\mu_1}j_{\nu_1}}\cdots \omega^{i_{\mu_r}j_{\nu_r}}\cr
 &{\pdr^r f^I\over \pdr \eta^{i_{\mu_1}
     a_1}_{b_1}\cdots \pdr \eta^{i_{\mu_r} a_r}_{b_r}}
{\pdr^r f^J\over \pdr \eta^{j_{\nu_1}
     b_1}_{a_1}\cdots \pdr \eta^{j_{\nu_r} b_r}_{a_r}}.\cr
}\eeqs
We can, using the symmetry of the derivatives, and relabelling of
indices, always bring the indices in the first derivative factor to
the order $\mu_1<\mu_2\cdots <\mu_r$. However, once this is done,
there is no reason  that the indices
$\nu_1,\nu_2\cdots \nu_r$ are in any particular  order.
This is because the contraction of the color indices links $\mu_k$ to
$\nu_k$.
Thus,
the general term in the series will involve quite complicated  ways
of contracting the color indices.

In the large $N_c$  limit, however,  the leading term will have the
largest number of traces. This will happen   when the
$\nu$ indices are in {\it decreasing }
order:$\{\nu_1>\nu_2\cdots > \nu_r\}$. This is the  term that involves a
product of $r$ Wilson  loops, so that
\beqs{
f^I*f^J&=f^If^J+\sum_{r=1}^{\infty}
\sum_{\matrix{\mu_1<\mu_2\cdots<\mu_r\cr
                              \nu_1>\nu_2\cdots > \nu_r\cr}}
     \bigg(-{i\hbar\over 2}\bigg)^r
\omega^{i_{\mu_1}j_{\nu_1}}\cdots \omega^{i_{\mu_r}j_{\nu_r}}\cr
&f^{I(\mu_1,\mu_2)J(\nu_2,\nu_1)}f^{I(\mu_2,\mu_3)J(\nu_3,\nu_2)}\cr
&\cdots f^{I(\mu_r,\mu_1)J(\nu_1,\nu_r)}+\cdots .
}\eeqs
Here $I(\mu_1,\mu_2)J(\nu_2,\nu_1)$ for example is the loop
$i_{\mu_1+1}i_{\mu_1+2}\cdots i_{\mu_2-1}
         j_{\nu_2+1}\cdots j_{\nu_1-1}
               $.\hfill\break

Now, as the large $N_c$ limit is  taken, we usually have to multiply
physical quantities by some  $N_c$--dependent factor, in order that
the limit  be well--defined. The proper normalization of $f^I$, for
example is not obvious. We propose that $f^I$ be normalized such that
its
vacuum expectation value remains finite in the limit. Now the vacuum
expectation value depends on the choice  of the Hamiltonian, which we
have not made yet. The simplest case would be a quadratic
function,$H=g_{ij}\tr\eta^i\eta^j$.
(This corresponds to having a free theory; if the coupling constants
in the interacting case
are scaled properly by powers of $N_c$ as well, the counting rules
 in powers of $N_c$  will not be affected.)
 The vacuum expectation value
of a product of two $\eta$'s is then
$<\eta^{ia}_b\eta^{jc}_d>=\eps^{ij}\delta^a_d\de^c_b$. (Here
$\eps^{ij}$ is a tensor built from $g_{ij}$ and $\omega_{ij}$ whose
explicit form is not necessary.)  The vacuum expectation value of
the product of an odd number of $\eta$'s will vanish; for an  even
number of $\eta$'s it is given by the Wick  formula.A short
calculation will show
that, for even $m$, the expectation value of $f^I$ is of order
$N_c^{{m\over 2}+1}$.   This is independent of the particular form of
the hamiltonian.

Thus we
define the normalized functions
\beq
     \tilde f^I={1\over N_c^{{m\over 2}+1} } f^I.
\eeq
Now one can check that
\beqs{
\tilde f^I*\tilde f^J&=\tilde f^I\tilde f^J+{1\over N_c^2}\sum_{r=1}^{\infty}
\sum_{\matrix{\mu_1<\mu_2\cdots<\mu_r\cr
                              \nu_1>\nu_2\cdots > \nu_r\cr}}
     \bigg(-{i\hbar\over 2}\bigg)^r
\omega^{i_{\mu_1}j_{\nu_1}}\cdots \omega^{i_{\mu_r}j_{\nu_r}}\cr
&\tilde f^{I(\mu_1,\mu_2)J(\nu_2,\nu_1)}
               \tilde f^{I(\mu_2,\mu_3)J(\nu_3,\nu_2)}\cr
&\cdots \tilde f^{I(\mu_r,\mu_1)J(\nu_1,\nu_r)}+ O({1\over N_c^3}) .
}\eeqs
Thus the pointwise product is again the leading contribution in the
large $N_c$ limit.

 The commutator  of the $\tilde f^I$'s
is of order ${1\over N_c^2}$.
This is consistent with the idea that the large $N_c$ limit is a
classical theory, with ${1\over N_c^2}$ measuring the size of the
quantum corrections:  analogous to the  $\hbar$ of  the conventional
classical limit.Indeed, the Poisson algebra of the large $N_c $ limit
 of (regularized) Yang-Mills theory is,
\beqs{
\{\tilde f^I ,
\tilde f^J\}_*
&=2i\sum_{r=1,\;{\rm odd}\;}^{\infty}
\sum_{\matrix{\mu_1<\mu_2\cdots<\mu_r\cr
                              \nu_1>\nu_2\cdots > \nu_r\cr}}
     \bigg(-{i\hbar\over 2}\bigg)^r
\omega^{i_{\mu_1}j_{\nu_1}}\cdots \omega^{i_{\mu_r}j_{\nu_r}}\cr
&\tilde f^{I(\mu_1,\mu_2)J(\nu_2,\nu_1)}
               \tilde f^{I(\mu_2,\mu_3)J(\nu_3,\nu_2)}\cr
&\cdots \tilde f^{I(\mu_r,\mu_1)J(\nu_1,\nu_r)}.
}\eeqs
The Poisson manifold defined by these relations
is the phase space of the large $N_c$ limit of Yang--Mills theory. It
is quite different from the phase space of the conventional classical
limit. (This is in  fact quite typical of such large $N_c$
limits.) We believe  that the above Poisson algebra plays a  fundamental
role in the physics of strongly interacting particles.

{\it Acknowldgement}\hfill\break

We thank Prof. S. Okubo for discussions. This work was supported in
part by the  US Department of Energy under the grant
DE-FG02-91ER40685.  S. G. R.  also thanks the Erwin Schr\"odinger
Institute, where this paper was written,  for hospitality.

\vfill\break

{\it References}\hfill\break

\loops.\ S. Mandelstam Phys. Rev. {\bf 175}, 1580 (1968); Phys. Rev.
{\bf D19}, 2391 (1979). Y. Makeenko  and A. A. Migdal, Nucl. Phys.
{\bf B188} 269 (1981).
 S. G. Rajeev, Ann. Phys. {\bf 173} 249
(1987); Phys. Lett. {\bf B212}, 203 (1988); Int. J. Mod. Phys.
A9:5583-5624(1994). A. Ashtekar, {\it Lectures on Non-perturbative
Canonical Gravity}, Singapore, World
Scientific(1991); A. Ashtekar and C. Isham, Class. Quan. Grav. {\bf 9} 1433
(1993).

\rajturgut.\  S. G. Rajeev, in  Proceedings of the 1991 Trieste Summer School
ed.
Gava, Sezgin and Narain, World Scientific (1991);
 S. G. Rajeev and O. T. Turgut, Int. J. Mod. Phys. A
{\bf 10} 2479 (1995); S. G. Rajeev and O.T.Turgut in {\it Proceedings of
MRST'95}.

\nullcone.\ R. J. Perry, Ann. Phys. {\bf 232} 116 (1994) and refereneces
therein.

\penrose.\ R. Penrose and W. Rindler, {\it Spinors and Space-time}, Cambridge
Univ. Press, 1986; R. Wald,
 {\it General Relativity}, Univ. of Chicago Press, 1984.

\free.\ J. P. Serre, {\it Lie Algebras and Lie Groups}, Springer-Verlag,
1992, N. Jacobson,{\it Basic Algebra, Vol-II}, 2nd Ed., W. H. Freeman, 1989;
for a more abstract presentation, P. M. Cohn,{\it Universal Algebra},
2nd Ed, D. Reidel Pub., 1981.

\connes\ A. Connes, {\it Non-commutative Geometry}, Academic Press, 1994.

\hormander.\ L. Hormander, Comm. Pure and Appl. Math. {\bf 32} 359 (1979);
G. B. Folland, {\it Harmonic Analysis in Phase space}, Princeton Univ. Press,
1989.

\bye